\documentclass[12pt]{iopart}

\newcommand{\En}{\mathcal{E}}
\newcommand{\Lz}{\mathcal{L}}
\newcommand{\Q}{\mathcal{Q}}
\newcommand{\C}{\mathcal{C}}
\newcommand{\K}{\mathcal{K}}
\newcommand{\sn}{\mathrm{sn}}
\newcommand{\am}{\mathrm{am}}

\usepackage{cite} 
\usepackage{setspace}
\usepackage{lscape}
\usepackage{import}
\usepackage{wrapfig}
\usepackage{multicol}
\usepackage{textcomp}
\usepackage{amsmath,accents}
\usepackage{amsfonts}
\usepackage{calc}
\usepackage{mathrsfs}
\usepackage{amssymb}
\usepackage{amsmath}
\usepackage{array}
\usepackage{color}
\usepackage{bm}
\usepackage{graphicx}
\usepackage{hyperref}
\usepackage{cleveref}
\usepackage{booktabs}
\usepackage{float}
\usepackage{mathtools}
\usepackage{caption}
\usepackage{subcaption}
\usepackage{tabularx}
\usepackage[scr=boondoxo,scrscaled=1.05]{mathalfa}
\usepackage[dvipsnames]{xcolor}
\usepackage[normalem]{ulem}
\usepackage{comment}
\usepackage{algorithm}
\usepackage{algpseudocode}

\begin{document}

\note[On the conversion of orbital angles for extreme mass ratio inspirals]{A note on the conversion of orbital angles for extreme mass ratio inspirals}

\author{Philip Lynch}
\address{Max Planck Institute for Gravitational Physics (Albert Einstein Institute), Potsdam-Golm, Germany}
\ead{philip.lynch@aei.mpg.de}

\author{Ollie Burke}
\address{School of Physics and Astronomy, University of Glasgow, Glasgow G12 8QQ, UK}

\address{Laboratoire des 2 Infinis -- Toulouse (L2IT-IN2P3), Universitè de Toulouse, CNRS, UPS, F-31062 Toulouse Cedex 9, France}

\vspace{10pt}

\begin{abstract}
We outline a practical scheme for converting between three commonly used sets of phases to describe the trajectories of extreme mass ratio inspirals; quasi-Keplerian angles, Mino time  action-angles, and Boyer-Lindquist time action-angles (as utilised by the \texttt{FastEMRIWaveform} package). Conversion between Boyer-Lindquist time action angles and quasi-Keplerian angles is essential for the construction of a source frame for adiabatic inspirals that can be related to the source frames used by other gravitational wave source modelling techniques. While converting from quasi-Keplerian angles to Boyer-Lindquist time action angles via Mino time  action-angles can be done analytically, the same does not hold for the converse, and so we make use of an efficient numerical root-finding method. We demonstrate the efficacy of our scheme by comparing two calculations for an eccentric and inclined geodesic orbit in Kerr spacetime using two different sets of orbital angles. We have made our implementations available in \texttt{Mathematica}, \texttt{C}, and \texttt{Python}.
\end{abstract}

%

\submitto{\CQG}
%
\maketitle
%
%

\section{Introduction}

With the advent of space-based gravitational wave (GW) detectors, such as the recently adopted Laser Interferometer Space Antenna (LISA), comes the need to produce highly accurate waveforms for gravitational wave (GW) sources in the millihertz frequency band \cite{LISAConsortiumWaveformWorkingGroup:2023arg,Colpi:2024xhw}. 
One major source of interest are systems consisting of a massive black hole (MBH) primary of mass $m_1 \sim 10^{5} - 10^7 M_\odot$ and a compact object (CO) secondary, such as a stellar mass black hole or neutron star, of mass $m_2 \sim 1- 10^2 M_\odot$ resulting in a binary with mass ratio $q = m_2 / m_1 \sim  10^{-3} - 10^{-7}$. The binary slowly loses energy and angular momentum due to the emission of GWs, which causes the secondary to eventually spiral into the primary. The result of this is a GW signal that will be present in the sensitivity band of LISA for months to years. These binary systems are known as extreme mass ratio inspirals (EMRIs) and are expected to be a key GW source for the LISA space mission. The trajectory of the smaller body will be highly relativistic, long-lasting and exceedingly complicated with respect to more comparable mass systems (see ~\cite{mancieri2024hangingcliffemriformation} and references within). Astrophysical formation channels suggest that the orbits will be both moderately eccentric and inclined with respect to the equatorial plane of the rotating primary black hole \cite{Gair:2004iv}. The result of these orbits is a phenomenologically rich waveform \cite{Drasco:2005kz} with features that strongly depend on the central black hole's parameters -- mass and Kerr rotation parameter. If modelled correctly, EMRI observations using LISA would provide unrivalled precision measurements on the astrophysical parameters that govern the system and our most stringent tests on Einstein's theory of General Relativity thus far~\cite{Amaro-Seoane:2007osp,Gair:2010iv,Babak:2017tow, speri2024probingfundamentalphysicsextreme, Burke_2020}.

One can approximate the system as a test particle following a bound, timelike geodesic in a black hole spacetime, such as Schwarzschild (non-rotating) or Kerr (rotating). To account for the emission of GWs, one can describe change of energy, angular momentum and Carter constant of the binary using so-called ``flux-balance formulae" \cite{Sago:2005fn,Isoyama:2018sib} in what is known as the adiabatic approximation (0PA) \cite{Hughes:1999bq, Hughes:2001jr,Glampedakis:2002ya,Drasco:2005kz,Hughes:2005qb,Hughes:2021exa,Isoyama:2021jjd, Nasipak:2023kuf}. Going beyond this level of accuracy requires additional knowledge of the local force acting on the test particle, which causes deviations from geodesic motion due to the COs own gravitational field, known as the gravitational self-force (GSF)~(see reviews~\cite{Barack_2018,Pound_2021} and and references within). The dynamics of the CO in the presence of the massive black hole can be accurately described using the two-timescale procedure~\cite{Hinderer:2008dm, Miller:2020bft}, where both the action and angle variables are presented as a perturbative expansion in the small mass-ratio $q$. It has been shown that in order to accurately track the phase evolution of a true vacuum GR based EMRI waveform to sub-radian accuracy, we require knowledge of the dissipative piece of the second-order in mass ratio GSF \cite{Pound:2012nt, Pound:2019lzj,Warburton:2021kwk, Wardell:2021fyy} It is thus both necessary, and sufficient to have these expansions known and correct to not only adiabatic (0PA) but post-adiabatic (1PA) order~\cite{Burke:2023lno}.  

One method for constructing the inspiral trajectory from the GSF assumes that the inspiral can be modelled as a series of instantaneously tangent geodesic orbits. This is known as the method of osculating geodesics and is described further in~\cite{Pound:2007th, Gair:2010iv}. This method is often too slow to be used on its own due to the need to accurately track $\mathcal{O}(q^{-1})$ orbital oscillations. Thus this method is coupled with averaging procedures such as near-identity (averaging) transformations (NITs) \cite{VanDeMeent:2018cgn,McCart:2021upc, Lynch:2021ogr,Lynch:2022zov,Lynch:2023gpu,Drummond:2023wqc,Lynch:2024ohd} or a two-timescale expansion \cite{Miller:2020bft, Hinderer:2008dm,Lynch:2023gpu} to accelerate the trajectory calculation. This all relies on a robust understanding of the leading-order, geodesic motion of the system.

Eccentric and inclined geodesics within the Kerr spacetime are tri-periodic and simultaneously display radial, polar, and azimuthal motion \cite{Drasco:2003ky}. Many works have thoroughly explored and described these orbits, but the choice of orbital phases used to describe the motion are not unique. One popular choice takes inspiration from Keplerian mechanics. The ``eccentric anomaly", $\psi$, was first used to describe eccentric orbits in Schwarzschild spacetimes~\cite{Dawrin1961} and a polar counterpart, $\chi$, was later constructed for circular and inclined orbits in Kerr spacetimes~\cite{Hughes:1999bq}. These phases have the advantage of having very simple relations to the radial and polar Boyer-Lindquist (BL) coordinates of the secondary, resulting in their widespread use to describe inspiral trajectories \cite{Pound:2007th,Gair:2010iv,Warburton:2011fk,Warburton:2017sxk,Osburn:2015duj,Drummond:2023loz}. However, the evolution equations for these phases do not have known closed form solutions for geodesic orbits and have to be solved numerically. For this work, we refer to this set along with the azimuthal BL coordinate $\phi$ as the ``quasi-Keplerian" (QK) angles.

Another choice was made possible by using (Carter-) Mino time (MT) $\lambda$ which decouples the radial and polar geodesic equations \cite{Mino:2003yg}, allowing for analytic expressions in terms of Jacobi-elliptic integrals for the radial and polar BL coordinates in terms of Mino Time action-angles (MTAAs) \cite{Drasco:2003ky,Fujita:2009bp,Kerachian:2023oiw}. The resulting expressions for the coordinates are more complicated than the QK ones. However, the rate of change of these phases with respect to Mino time are exactly the Mino time fundamental frequencies $\Upsilon_i$. Since these frequencies are constants of motion they are preferable to the QK angles when constructing NIT equations of motion for generic Kerr orbits \cite{Lynch:2021ogr,Lynch:2022zov,Lynch:2023gpu,Lynch:2024ohd}.

Finally, one can also describe the motion in terms of Boyer-Lindquist time action-angles (BLAAs) or ``true anomaly'' angles. These angles have no known analytic relation to the BL coordinates of the secondary. However, their rate of change of these phases with respect to BL time $t$ is exactly the BL time fundamental frequencies $\Omega_i$ \cite{Schmidt:2002qk}. The BL time coordinate is easily related to the time coordinate of the detector. As such, these AAs are very convenient for constructing efficient adiabatic EMRI waveforms via the multi-voice decomposition \cite{Hughes:2021exa} and are the angles utilised in the \texttt{FastEMRIWaveforms} (FEW) package \cite{Chua:2020stf,Katz:2021yft,Speri:2023jte,Chapman-Bird:2025xtd} as well as in some NIT inspiral calculations \cite{Pound:2019lzj,Lynch:2023gpu,Drummond:2023wqc}. Furthermore, these phases underlie the multi-scale analysis of the Einstein field equations needed to obtain 1PA results \cite{Miller:2020bft}.


Being able to convert between different sets of orbital angles is essential for comparing and validating trajectory calculations that use different parameterisations. In particular, a practical method for relating the BLAAs to the BL coordinates of the secondary enables the visualisation of 0PA inspirals. More importantly, it allows for the construction of a consistent source frame at 0PA order, which can then be translated into alternative frame conventions used in other LISA source modelling frameworks \cite{lisa_rosetta_stone_2025}.

Conversions between all these angles are known, but only in the form of Fourier decompositions that are not very practical to work with \cite{Drasco:2003ky}. An analytic transformation between QK angles and BL time AAs has been found for eccentric Schwarzschild orbits but relies on a small eccentricity expansion \cite{Witzany:2022vck}. More tractable relationships between the MTAAs and BLAAs for generic inspirals have been found \cite{Pound_2021} and implemented \cite{Lynch:2023gpu,Drummond:2023wqc}. 

\begin{figure*}
    \includegraphics[trim={0cm 6cm 0cm 6cm},clip,width = \textwidth]{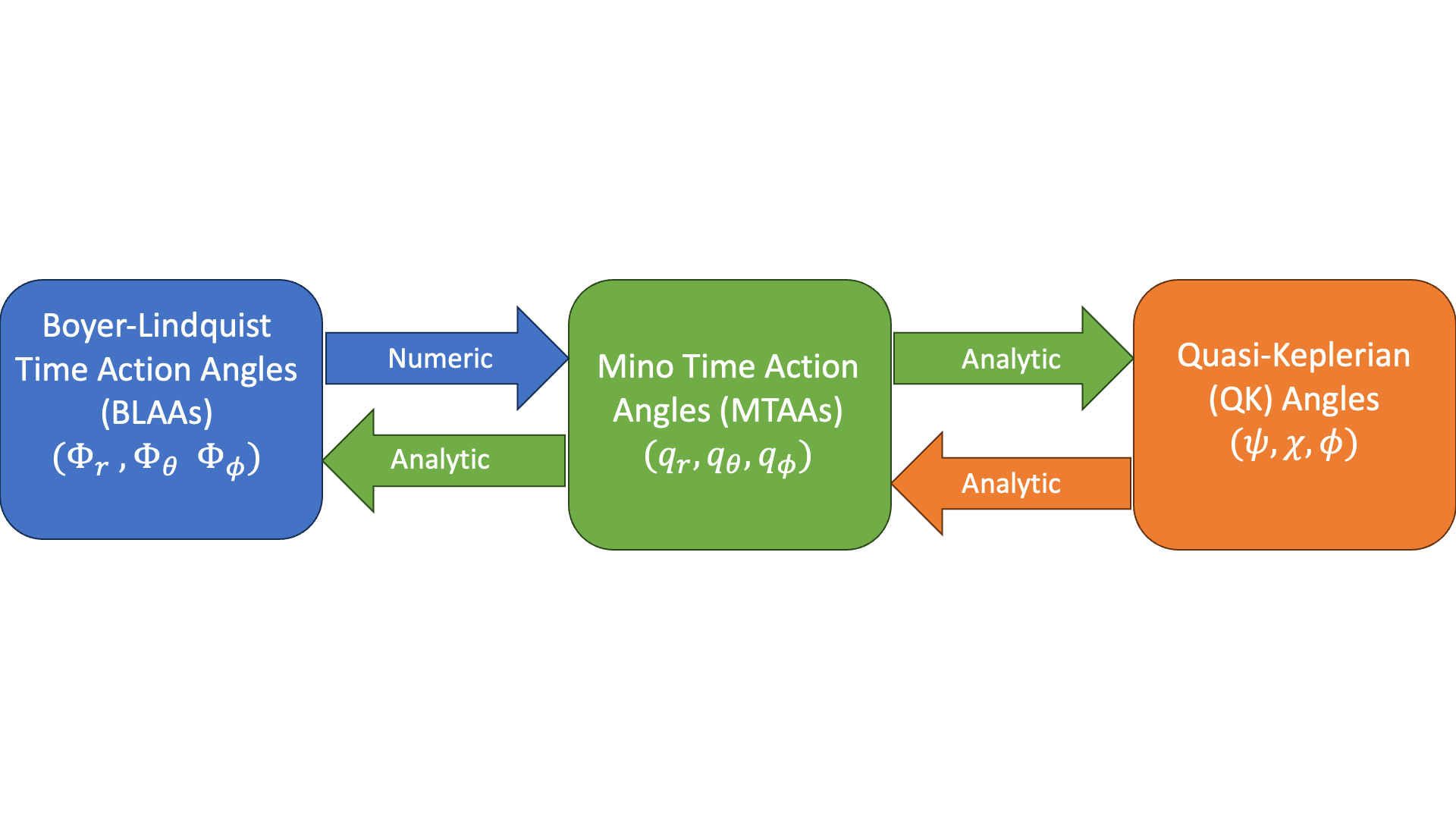}
    \caption{Schematic description of the analytical and numerical conversions between different sets of orbital angles as described in this note.}
    \label{fig:Flow_Chart}
\end{figure*}

This note looks to complete the picture by describing practical conversions between these three sets of orbital phases for generic Kerr orbits, as displayed in Fig.~\ref{fig:Flow_Chart}. We begin by outlining how one can analytically convert back and forth between QK angles and MTAAs in Sec.~\ref{section:QK_to_Mino_AAs}. We then show how one can find an analytic conversion from the MTAAs to the BLAAs, and how one can leverage a numerical root-finding scheme to  invert this relationship in Sec.~\ref{section:Mino_to_BL_AAs}. With this in hand, it is then straightforward to convert from BLAAs to QK angles via the MTAAs, as described in Sec.~\ref{section:QK_to_BL_AAs}. 
We demonstrate our methods in Sec.~\ref{section:Results} by comparing between two different calculations for the same eccentric and inclined Kerr geodesic. We have made these transformations public as part of the \texttt{KerrOrbitalAngleConversion} Github repository \cite{KerrAngleConversions} and they will soon be implemented as part of the \texttt{FEW} package and the \texttt{KerrGeodesics} Mathematica package \cite{KerrGeodesics}.

\section{Geodesics in Kerr spacetime}

\subsection{Constants of Motion}
We denote the mass of the primary black hole by $m_1$ and parameterise the black hole's rotation by $a = |J|/m_1$ where $J$ is the angular momentum of the black hole.	The Kerr metric can then be written in modified BL coordinates, $x^\alpha = (t,r,z=\cos\theta,\phi)$, as
	\begin{equation}\label{eq:metric}
		\begin{split}
			ds^2 = & - \left(1 - \frac{2 m_1 r}{\Sigma} \right) dt^2 + \frac{\Sigma}{\Delta} dr^2 + \frac{\Sigma}{1-z^2} dz^2 \\                 & + \frac{1-z^2}{\Sigma} (2a^2 m_1 r (1 - z^2) + \Sigma \varpi^2) d\phi^2 - \frac{4 m_1 a r (1-z^2)}{\Sigma}dt d\phi
		\end{split}
	\end{equation}
	where
	\begin{subequations}\label{eq:DeltaSigmaOmega}
		\begin{gather}
			\Delta(r) := r^2 + a^2 - 2 m_1 r,   \quad   \Sigma(r,z) := r^2 + a^2 z^2,   \quad   \varpi(r) := \sqrt{r^2 + a^2},    \tag{\theequation a-c}
		\end{gather}
	\end{subequations}
		In the absence of any perturbing force, the secondary will follow a geodesic, i.e., $ u^\beta\nabla_\beta u^\alpha = 0$, where $u^\alpha$ denotes the four-velocity of the secondary.
	The symmetries of the Kerr spacetime allow for the identification of integrals of motion $\vec{\mathcal{P}} = ( \En, \Lz, \C)$ given by
	\begin{subequations}\label{eq:ELK}
		\begin{gather}
			\En = - u_t,   \quad   \Lz = u_\phi,   \quad   \C = \K^{\alpha \beta} u_\alpha u_\beta,    \tag{\theequation a-c}
		\end{gather}
	\end{subequations}
	where $\En$ is the orbital energy per unit rest mass $m_2$, $\Lz$ is the z-component of the angular momentum divided by $m_2$, $\K^{\alpha \beta}$ is the Killing tensor, and $\C$ is the Carter constant divided by $m_2^2$ \cite{Carter:1968rr}.
	This definition of the Carter constant is related to another common definition of the Carter Constant via $\Q = \C - (\Lz - a \En)^2$.
	Using these integrals of motion, one can express the geodesic equations in first order, decoupled form \cite{Drasco:2003ky}:
\begin{subequations}\label{eq:Geodesic_eqs}
	\begin{align}
		\begin{split}\label{eq:Vr}
			\left( \frac{d r}{d\lambda} \right)^2 &= \mathcal{B} ^2 - \Delta\left(r^2+ \C \right) \\
			&= (1-\En^2)(r_a-r)(r-r_p)(r-r_3)(r-r_4) \coloneqq V_r,
		\end{split}\\
		\begin{split} \label{eq:Vz}
			\left( \frac{d z}{d\lambda} \right)^2 &= \Q - z^2 \left(\beta (1-z^2) + \Lz^2 + \Q \right) \\
			&= (z^2-z_-^2)\left(\beta z^2 - z_+^2 \right) \coloneqq V_z,
		\end{split}\\
		\begin{split}\label{eq:Geodesic_t}
			\frac{d t}{d\lambda} &= \frac{\varpi^2}{\Delta} \mathcal{B} - a^2 \En(1-z^2)+ a \Lz 
			\coloneqq f_t,
		\end{split}\\
		\begin{split} \label{eq:Geodesic_phi}
			\frac{d \phi}{d\lambda} &= \frac{a}{\Delta} \mathcal{B} + \frac{\Lz}{1-z^2} - a \En   \coloneqq f_\phi,
		\end{split}
	\end{align}
\end{subequations}
	where $\mathcal{B} = \En \varpi^2 - a \Lz $, $\beta = a^2 (1-\En^2) $. The roots of the radial potential $V_r$ are denoted $r_a > r_p > r_3 > r_4$ where the radial motion of the secondary is confined between the radius of apastron $r_a$ and the radius of periastron $r_p$. Similarly, the roots of the polar potential $V_z$ are denoted  $z_+ > z_-$ where the polar motion of the secondary is bounded between $z_-$ and $-z_-$. Finally, note that we are making use of Mino(-Carter) time $\lambda$ so that we may decouple the radial and polar equations \cite{Mino:2003yg}. 
	This time is related to the proper time of the particle, $\tau$, by $d\tau = \Sigma d \lambda$.

\begin{figure}
		\centering
		\includegraphics[width = \textwidth]{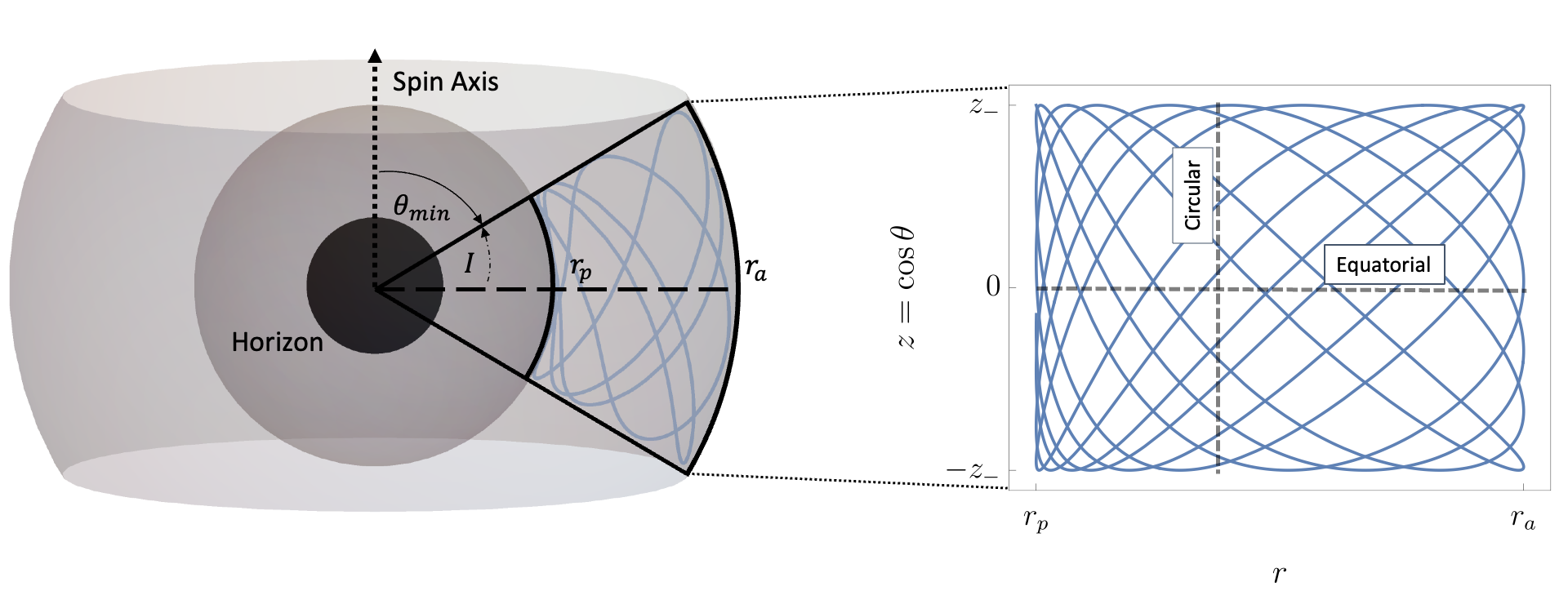}
		\caption{A illustration of the radial and polar motion present for generic Kerr orbits. The graph shows how the motion covers  a torus which is bounded by the radial roots $r_a$ and $r_p$ and the value of the polar root $z_- = \cos \theta_\text{min}$.}
		\label{fig:Kerr_Geodesics}
	\end{figure}

Rather than parameterise an orbit by the set $\vec{\mathcal{P}} = (\En, \Lz, \C)$ it is useful to instead use more geometric, quasi-Keplerian constants $\vec{P} = (p,e,x_I)$.
	Here $p$ is the semi-latus rectum and $e$ is the orbital eccentricity which are both related to $r_a$ and $r_p$ via:
	\begin{subequations}\label{eq:primary_roots_inverse}
		\begin{gather}
			p = \frac{2 r_a r_p}{r_a + r_p} \quad \text{and}   \quad e = \frac{r_a-r_p}{r_a+r_p}.    \tag{\theequation a-b}
		\end{gather}
	\end{subequations}
	We also use $x_I$ which is a measure of the orbital inclination given by
	\begin{equation}\label{eq:x_def}
		x_I = \cos I = \sqrt{1-z_-^2},
	\end{equation}
    where $x_I=1$ corresponds to pro-grade equatorial motion and  $x_I=-1$ corresponds to retrograde equatorial motion.
	The inclination angle $I$ is related to $\theta_{\text{min}}$ (the minimum value of $\theta$ measured with respect to the primary's spin axis) by $I = \pi / 2 - \text{sgn}(\Lz) \theta_{\text{min}}$.
	These angles and their relationship to each other are illustrated in Fig.~\ref{fig:Kerr_Geodesics}\footnote{The 5PN Augmented Analytic Kludge model in FEW uses a different definition of inclination given by $Y = \cos \iota = \Lz(\vec{P}) / \sqrt{\Lz(\vec{P})^2 + Q(\vec{P})}$. The conversion from $x_{I}$ to $Y$ is trivial and a function to numerically convert from $Y$ to $x_I$ is provided in the FEW package~\cite{Katz:2021yft}.}.
	
	We opt to use $\vec{P}$ as they are easily related to the radial and polar roots via
	\begin{subequations}\label{eq:primary_roots}
		\begin{gather}
			r_a = \frac{p }{1-e},   \quad  r_p = \frac{p }{1+e},   \quad   z_- = \sqrt{1-x_I^2}.    \tag{\theequation a-c}
		\end{gather}
	\end{subequations}
    One can also determine expressions for $\vec{\mathcal{P}}$ in terms of $(r_a,r_p,z_-)$ and thus in terms of $\vec{P}$ as was done in Ref.~\cite{Schmidt:2002qk}.
 With these expressions in hand, one can determine the remaining roots of the radial and polar potentials via \cite{Fujita:2009bp,vandeMeent:2019cam}
	\begin{subequations}\label{eq:secondary_roots}
		\begin{align}
			\begin{split}
				r_3 &= \frac{m_1}{1-\En^2} - \frac{r_a+r_p}{2} + \sqrt{\left( \frac{r_a+r_p}{2} -\frac{m_1}{1-\En^2} \right)^2 - \frac{a^2\Q}{r_a r_p(1-\En^2)}}
			\end{split}\\
			\begin{split}
				r_4 &=\frac{a^2\Q}{r_a r_p r_3(1-\En^2)},
			\end{split}\\
			\begin{split}
				z_+ &= \sqrt{\beta + \frac{\Lz^2}{x_I^2}}.
			\end{split}
		\end{align}
	\end{subequations}
	 As such, given values for $\vec{P}$, we can determine all of the other constants that appear in the geodesic equations of motion.

\subsection{Quasi-Keplerian parameterisation}
We note that it is also common in the literature \cite{Drasco:2003ky, Gair:2010iv} to express the radial and polar motion in terms of quasi-Keplerian angles \cite{Dawrin1961, Hughes:1999bq}\footnote{These are not the only Keplerian inspired angles that exist in the literature. See for example Ref.\cite{Piovano:2024yks} for their construction and conversions to other sets of angles.}, $\psi$ and $\chi$, via
	\begin{subequations}\label{eq:Keplerian_Angles}
		\begin{gather}
			r(\psi) = \frac{p}{ 1 + e \cos\psi} \quad \text{and}  \quad z(\chi) = z_- \cos \chi.    \tag{\theequation a-b}
		\end{gather}
	\end{subequations}
	With this parameterisation, the evolution equations for $\psi$ and $\chi$ depend on $\psi$ and $\chi$ respectively i.e.,
	\begin{subequations} \label{eq:Keplerian_Frequencies}
		\begin{align}
			\begin{split}
			\frac{d \psi}{d \lambda} = \frac{1}{1-e^2} \sqrt{(1-\En^2) \left((p - p_3) - e(p+p_3 \cos \psi)\right) \left((p - p_4) + e(p-p_4 \cos \psi)\right)} \coloneqq f_r(\psi), 
			\end{split}\\
		\begin{split}
			\frac{d \chi}{d \lambda} =  \sqrt{(z_+^2 - \beta z_-^2 \cos^2 \chi)} \coloneqq f_\theta (\chi),  
		\end{split}
		\end{align}
	\end{subequations}
	where $p_3 = r_3(1-e)$ and $p_4 = r_4(1+e)$ \cite{Drasco:2003ky}. These evolution equations currently have no known closed form solution and so are typically solved numerically.
\subsection{Mino time action angle parametrisation}
To obtain analytical solutions to the geodesic motion, one can make use of Mino-time action angles $\vec{q} = (q_t,q_r,q_\theta,q_\phi)$. The geodesic equations can rewritten in terms of $\vec{q}$ and the orbital elements $\vec{P}$ to read
	\begin{subequations}\label{eq:GeodesicEqsPandq}
		\begin{gather}
			\frac{d P_j}{d \lambda} = 0\quad \text{and}  \quad \frac{d q_i}{d \lambda} =  \Upsilon_i (\vec{P}),   \tag{\theequation a-b}
		\end{gather}
	\end{subequations}
	where $\Upsilon_i (\vec{P})$ are the Mino time fundamental frequencies which have analytic expressions first derived in Ref. \cite{Fujita:2009bp}.
 From the geodesic equations of motion, the solutions for $\vec{P}(\lambda)$ and $\vec{q}_i(\lambda)$ are simply given by
	\begin{subequations}\label{eq:GeodesicEqsPandq_solutions}
		\begin{gather}
			P_j(\lambda) = P_{j,0} \quad \text{and}  \quad  q_i=  \Upsilon_i (\vec{P}_0) \lambda + q_{i,0},  \tag{\theequation a-b}
		\end{gather}
	\end{subequations}
	where $P_{j,0}$ and $q_{i,0}$ are the initial values at $\lambda = 0$.

    Using the expressions derived in Ref.~\cite{Fujita:2009bp} and then simplified in Ref.~\cite{vandeMeent:2019cam}, one can express the radial and polar BL coordinates of a test particle under geodesic motion via
	\begin{subequations}
		\begin{align}
		\begin{split} \label{eq:r_analytic}
			r(q_r) &= \frac{r_3 (r_a - r_p)\sn^2\left(\frac{K(k_r)}{\pi} q_r | k_r\right) - r_p(r_a-r_3)}
			{(r_a - r_p) \sn^2\left(\frac{K(k_r)}{\pi} q_r | k_r\right) - (r_a - r_3)},
		\end{split}\\
		\begin{split} \label{eq:z_analytic}
		z(q_\theta) &= z_-\sn\left(K(k_z)\frac{2(q_\theta+\frac{\pi}{2})}{\pi} | k_z\right),
		\end{split}
	\end{align}
	\end{subequations}
    where
	\begin{subequations}
		\begin{gather}
			k_r= \frac{(r_a - r_p)(r_3-r_4)}{(r_a - r_3)(r_p - r_4)},  \quad \text{and} \quad  k_z= \beta \frac{z_-^2}{z_+^2}.    \tag{\theequation a-b}
		\end{gather}
	\end{subequations}
 Here, $K$ is complete elliptic integral of the first kind given by $K(m) = F(\pi/2|m)$ and $F$ is the incomplete elliptic integral of the first kind given by
	\begin{equation} \label{eq:Elliptic_Integral_First_Kind}
		F(\phi|m) \coloneqq \int_0^\phi \frac{d\theta}{\sqrt{1 - m \sin^2 \theta}}.
	\end{equation}
 The inverse of this function is the Jacobi amplitude $\am$ i.e. if $u = F(\phi | m)$, then $\phi = \am(u|m)$.
 Thus Jacobi elliptic sine function $\sn$  given by $\sn(u|m) = \sin (\am(u|m))$.
	Furthermore, we have adopted the convention that $q_r = 0$ corresponds to the periapsis of the radial motion (i.e. $r = r_p$), and $q_\theta = 0$ corresponds to the maximum value of the polar BL coordinate $z$ (i.e. $z = z_-$).

 Ref.\cite{Fujita:2009bp} also derived analytic solutions for the Boyer-Lindquist time and azimuthal geodesic equations,  \eqref{eq:Geodesic_t} and \eqref{eq:Geodesic_phi}. These solutions can be written schematically as:
	\begin{subequations}
		\begin{align}
			\begin{split}	\label{eq:t_analytic}
				t(\lambda) &= q_t+ \delta t (q_r,q_\theta) =  \Upsilon_t\lambda + q_{t,0} + t_r(q_r) + t_\theta(q_\theta) ,
			\end{split} \\
			\begin{split}	\label{eq:phi_analytic}
				\phi(\lambda) &= q_\phi + \delta \phi(q_r,q_\theta) = \Upsilon_\phi\lambda + q_{\phi,0} + \phi_r(q_r) + \phi_\theta(q_\theta),
			\end{split}
		\end{align}
	\end{subequations}
where we use $\delta f$ to denote that a function is $2 \pi$ periodic in its arguments and has an orbit average of zero, and $t_r \backslash \phi_r$ and $t_\theta \backslash \phi_\theta$ are oscillatory functions that depend solely on the radial and polar phases respectively.
Their expressions can be found in full in \ref{section:Analytic_expressions_for_t_and_phi}. For this work, we choose the convention that $q_{t,0} = 0$.

\subsection{Boyer-Lindquist time action angle parameterisation}
We note that the Mino time fundamental frequencies $(\Upsilon_t,\Upsilon_r,\Upsilon_z,\Upsilon_\phi )$ do not have a physical interpretation by themselves. However, the frequencies observed at infinity, i.e., the BL time fundamental frequencies $\vec{\Omega} = (\Omega_r, \Omega_\theta,\Omega_\phi)$, are given by
	\begin{subequations} \label{eq:Boyer-Lindquist_Frequencies}
	\begin{gather}
		\Omega_r = \frac{\Upsilon_r}{\Upsilon_t}, \quad \Omega_z = \frac{\Upsilon_z}{\Upsilon_t}, \quad \Omega_\phi = \frac{\Upsilon_\phi}{\Upsilon_t}.  \tag{\theequation a-c}
	\end{gather}
	\end{subequations}
 From these, one can define a another set of action angles associated with the BL time coordinate that, for geodesics, follow the equations of motion:

 \begin{subequations}\label{eq:GeodesicEqsPandPhi}
		\begin{gather}
			\frac{d P_j}{d t} = 0\quad \text{and}  \quad \frac{d \Phi_i}{d t} =  \Omega_i (\vec{P}).   \tag{\theequation a-b}
		\end{gather}
	\end{subequations}
These have the solutions
\begin{subequations}\label{eq:GeodesicEqsPandPhi_solutions}
		\begin{gather}
			P_j(t) = P_{j,0} \quad \text{and}  \quad  \Phi_i=  \Omega_i (\vec{P}(0)) t + \Phi_{i,0},  \tag{\theequation a-b}
		\end{gather}
	\end{subequations}
	where $P_{j,0}$ and $\Phi_{i,0}$ are the initial values at $t = 0$.

\section{Angle Conversions} \label{section:Angle_Conversions}
\subsection{Quasi-Keplerian to Mino-time action angles} \label{section:QK_to_Mino_AAs}
We now seek to establish a practical method to relate the various different phases to one another.
We start by relating the radial QK angle $\psi$ to the radial MTAA $q_r$. Using \eqref{eq:Keplerian_Angles} and \eqref{eq:r_analytic}, we can equate the analytic solutions for the radial BL coordinate $r$, i.e., $r(\psi) = r(q_r)$ to obtain
\begin{subequations} \label{eq:Radial_Mino_to_QK}
    \begin{equation}\label{eq:psi_of_qr}
        \psi(q_r) = \arccos \left(  \frac{p - r(q_r)}{ e r(q_r)} \right)\,,
    \end{equation}
    \begin{equation}\label{eq:qr_of_psi}
        q_r(\psi) = \frac{\pi F\left(\arcsin  (k_\psi(\psi) ) | k_r \right)} {K(k_r)}\,.
    \end{equation}
\end{subequations}
Here $k_\psi = \sqrt{\frac{(r(\psi)-r_p)(r_a - r_3)}{(r(\psi) - r_3)(r_a-r_p)}}$. While this relationship appears to break down for circular orbits where $e = 0$, we show in \ref{section:Radial_Conv_Circular_Limit} that $\lim_{e \rightarrow 0} q_r = \psi$ and vice versa. 

Similarly, one can relate the polar QK angle $\chi$ to the polar MTAA $q_\theta$ by equating $z(\chi) = z(q_\theta)$, which gives us
\begin{subequations} \label{eq:Polar_Mino_to_QK}
    \begin{equation}\label{eq:chi_of_qtheta}
        \chi(q_\theta) = \arccos \left( \frac{z(q_\theta)}{z_-}\right),
    \end{equation}
    \begin{equation}\label{eq:qtheta_of_chi}
        q_\theta(\chi) = \frac{\pi}{2} - \frac{\pi F\left(\arcsin  (\cos \chi )| k_z \right)} {2 K(k_z)}\,.
    \end{equation}
\end{subequations}
This conversion appears to break down in the equatorial plane when $z_- = 0$. However, in \ref{section:Polar_Conv_Equatorial_Limit} we show that that $\lim_{z_- \rightarrow 0} q_\theta = \chi$ and vice versa.


Note that our conversions rely on the use of functions $\arccos$ and $\arcsin$ where most numerical implementations return values in the ranges $[0,\pi]$ and $[-\pi/2,\pi/2]$ respectively. This is at odds with our use-case where we treat these phases as monotonically increasing quantities. If $\varphi$ is the angle we wish to convert and $\tilde{\varphi}$ is the converted angle, then we can unwrap the converted angle using the following simple procedure: 
\begin{algorithm}
\begin{algorithmic}
\State $B \gets \varphi - \left( \varphi \mod 2 \pi \right)$ \Comment{We identify the ``bulk'' of the angle}
\State $\varphi \gets \varphi\mod 2 \pi $ \Comment{Then we take the angle modulo $2 \pi$}
\State $\tilde{\varphi} \gets \texttt{conversion}(\varphi)$ \Comment{We apply the conversion i.e. Eq.~\eqref{eq:Radial_Mino_to_QK} or Eq.~\eqref{eq:Polar_Mino_to_QK}}
\If{$\varphi \leq \pi$}
    \State $R \gets \tilde{\varphi}$ \Comment{We identify the ``remainder''}
\ElsIf{$\varphi> \pi$}
    \State $R = 2\pi -\tilde{\varphi}$
\EndIf
\State $\tilde{\varphi} \gets B + R$ \Comment{The unwrapped angle is the ``bulk" + the ``remainder''}
\end{algorithmic}
\end{algorithm}

With these conversions in place for the radial and polar phases, converting between the azimuthal phases $\phi$ and $q_\phi$ can be straightforwardly done using Eq.~\eqref{eq:phi_analytic}. Similarly, one can relate the BL time coordinate $t$ to $q_t$ using Eq.~\eqref{eq:t_analytic}.

\subsection{Mino-time action angles to Boyer-Lindquist action angles} \label{section:Mino_to_BL_AAs}
We now look to convert between the MTAAs $\vec{q} = \left(q_r,q_\theta,q_\phi \right) $ and the BLAAs $\vec{\Phi} = \left(\Phi_r,\Phi_\theta,\Phi_\phi \right)$. Without loss of generality, we assume that at $\lambda = 0$ the orbit starts at $r = r_p$ and $\theta = \theta_-$ and thus choose initial conditions such that $q_{i,0} = \Phi_{i,0} = 0$ \footnote{One can eventually derive the same result using arbitrary initial conditions as was done in an earlier draft of this manuscript, but the derivation becomes significantly longer.}. Inserting Eq.~\eqref{eq:t_analytic} into Eq.~(\ref{eq:GeodesicEqsPandPhi_solutions}b) and recalling Eqs.~\eqref{eq:Boyer-Lindquist_Frequencies} and Eq.~(\ref{eq:GeodesicEqsPandq_solutions},b), one obtains
\begin{equation} \label{eq:Mino_to_BL}
    \Phi_i = \Omega_i \left( \Upsilon_t \lambda+ \delta t  \right) = \Upsilon_i\lambda + \Omega_i \delta t = q_i + \Omega_i \delta t.
\end{equation} 
One can then use the above equation to determine the BLAA initial conditions $\Phi_{i,0}$ if given any other MTAA initial conditions $q_{i,0}$.

Unfortunately, there is no way to analytically invert this relationship to convert from $\Phi_r$ and $\Phi_\theta$ to $q_r$ and $q_\theta$ for two reasons. Firstly, the radial and polar equations in BL time are coupled. Secondly, even in the limits where these equations decouple, such as the circular orbit limit and the equatorial plane, we are left with a more complicated version of the Kepler equation which is known to have no closed form solution for its inverse except for infinite series expansions \cite{colwell1993solving}. The most practical approach is to chose our initial estimate to be $q_{r} \sim \Phi_r$ and $q_{\theta} \sim \Phi_\theta$ and then use an iterative root-finding method on the coupled system of Eqs.~\eqref{eq:Mino_to_BL}
to numerically find the solutions for $q_r(\Phi_r, \Phi_\theta)$ and $q_\theta(\Phi_r, \Phi_\theta)$. With these numerical solutions in hand, one can then freely convert between the two azimuthal phases using:
    \begin{equation}\label{eq:BL_to_Mino_Azimuthal}
        q_\phi =  \Phi_\phi - \Omega_\phi \delta t(q_r(\Phi_r, \Phi_\theta),q_\theta(\Phi_r, \Phi_\theta)).
    \end{equation}

\subsection{Quasi-Keplerian to Boyer-Lindquist action angles} \label{section:QK_to_BL_AAs}
Finally, if one wishes to convert from QK angles $(\psi,\chi,\phi)$ to BLAAs $(\Phi_r,\Phi_\theta,\Phi_\phi)$, one can do so analytically. First, one uses Eqs.~\eqref{eq:qr_of_psi}, \eqref{eq:qtheta_of_chi}, and \eqref{eq:phi_analytic} to obtain values for the MTAAs $(q_r,q_\theta,q_\phi)$. Using these, one can then use Eqs.~\eqref{eq:Mino_to_BL} to obtain values for $(\Phi_r,\Phi_\theta,\Phi_\phi)$.

As mentioned in the previous section, there is not an analytic conversion to go from $(\Phi_r,\Phi_\theta,\Phi_\phi)$ back to $(\psi,\chi,\phi)$. We opt to use a root finding method on Eqs.~\eqref{eq:Mino_to_BL}
to numerically find values for $q_r$ and $q_\theta$ and then use these values with Eq.~\eqref{eq:BL_to_Mino_Azimuthal} to obtain $q_\phi$. Finally we use Eqs.~\eqref{eq:psi_of_qr}, \eqref{eq:chi_of_qtheta}, and \eqref{eq:phi_analytic} to obtain our final values of $(\psi,\chi,\phi)$.

We have developed practical implementations of these schemes which can be used in Wolfram language (Mathematica), C, and Python (using Cython) and have been made available via the KerrOrbitalAngleConversion Github Repository \cite{KerrAngleConversions}. In Mathematica, we make use of the \texttt{FindRoot} function, while in C we use the root-finding scheme present in the GNU Scientific Library \cite{galassi2009gnu}.

\section{Results} \label{section:Results}

\begin{figure} 
    \centering
    \includegraphics[clip,width=0.9\textwidth]{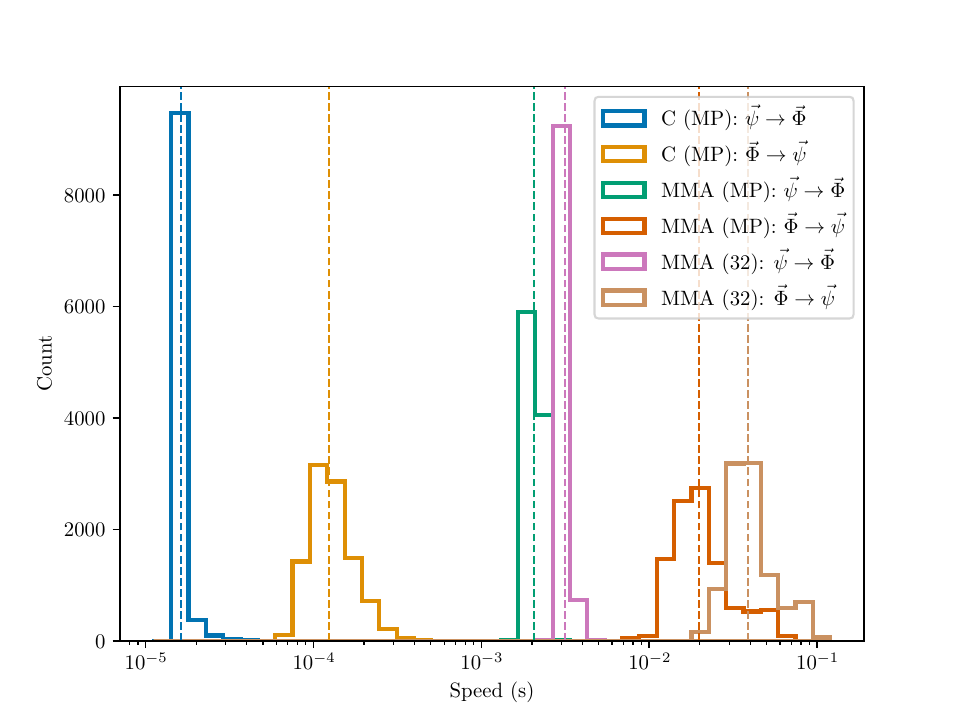}
    \caption{The timings of the transformations to and from QK angles $\vec{\psi}$ and BLAAs $\vec{\Phi}$ as performed on $10^4$ randomly drawn parameters for both our Cython (C) and Mathematica (MMA) implementations. For our MMA implementation, we also include timings when using a working precision of machine precision (MP) and 32 digits of precision.}
    \label{fig:Timings}
\end{figure}

To investigate the efficiency of our transformations, we draw a sample of $10^4$ values for $(a,p,e,x_I,\psi, \chi, \phi)$ and perform both a transformation from QK angles $\vec{\psi}$ to BLAAs $\vec{\Phi}$ and back. We report the timings for each transformation using both our Cython implementation and our Mathematica implementation in Fig.~\ref{fig:Timings} as recorded on an Mac M1 Max.  The transformation from QK angles to BLAA relies on the evaluation of elliptic integrals and takes a median of $1.63 \times 10^{-5}$ s or $2 \times 10^{-3}$ s in Cython and Mathematica respectively. The inverse transformation is expectedly slower, due to the need to numerically root-find, taking a median of $2.05 \times 10^{-4}$ s or $1.99 \times 10^{-2}$ s. One of the advantages of Mathematica is the ease of using arbitrary precision. We show that using 32 digits instead of machine precision numbers results more accurate transformations at the cost of slightly slower transformations of median $3.14 \times 10^{-3}$ s and $3.86 \times 10^{-2}$ s.

To demonstrate the accuracy of our Cython implementation, we will evolve a geodesic trajectory with orbital parameters $(a,p,e,x_I) = (0.9,7,0.5,0.75)$ and initial conditions $(\psi_0,\chi_0,\phi_0) = (2.7,1.8.0.9)$. We will evolve this system from $t = 0$ to $t = 3000 m_1$ using two independent methods. First, by using a 4th order Runge-Kutta scheme, we numerically solve for the quasi-Keplerian (QK) phases $(\psi(t),\chi(t),\phi(t))$ which obey

\begin{subequations} \label{eq:Quasi-Keplerian_EoM}
    \begin{gather}
			\frac{d \psi}{d t}  = \frac{f_r}{f_t},   \quad  \frac{d \chi}{d t}  = \frac{f_\theta}{f_t}, \quad \text{and}  \quad \frac{d \phi}{d t}  = \frac{f_\phi}{f_t},  \tag{\theequation a-c}
		\end{gather}
\end{subequations}
with forcing functions $(f_{t}, f_{\phi},f_{r}, f_{\theta})$ given by equations \eqref{eq:Geodesic_t}, \eqref{eq:Geodesic_phi} and sub-equations \eqref{eq:Keplerian_Frequencies} respectively.
Secondly, we solve for Boyer-Lindquist action angles (AA) $(\Phi_r(t), \Phi_\theta(t), \Phi_\phi(t))$ analytically via Eqs.~\eqref{eq:GeodesicEqsPandPhi_solutions}. 
We then use the conversions between the QK and BL time AA angles described in Sec.~\ref{section:QK_to_BL_AAs} to compare the solutions for the two sets of angles and access the accuracy of the conversions. 

We find that performing a single conversion from the QK initial conditions to the AA initial conditions with the absolute error tolerance of the root-finder set to $10^{-10}$, takes $\sim 0.2$ ms, while the inverse transformation only takes $\sim 0.07$ ms. Comparing the inverse to the original shows that that the error is consistent with the tolerance of the root-finder. 

\begin{figure}
	\begin{subfigure}[b]{0.32\textwidth}
		\centering
		\includegraphics[clip,width=\textwidth]{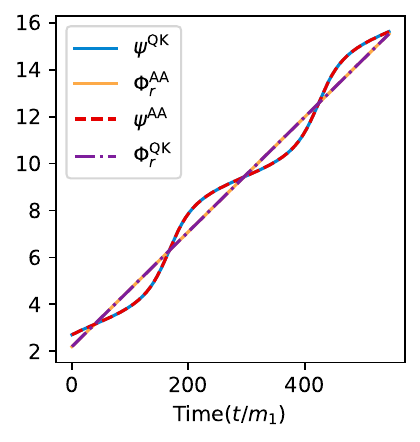}
		\caption{Radial Phases}
		\label{fig:Radial_Phases}
	\end{subfigure}
	\hfill
	\begin{subfigure}[b]{0.32\textwidth}
		\centering
		\includegraphics[width=\textwidth]{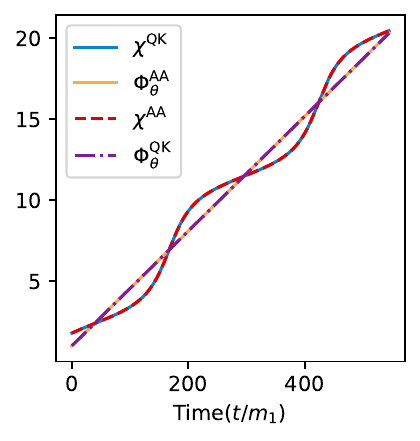}
		\caption{Polar Phases}
		\label{fig:Polar_Phases}
	\end{subfigure}
	\hfill
	\begin{subfigure}[b]{0.32\textwidth}
		\centering
		\includegraphics[clip,width=\textwidth]{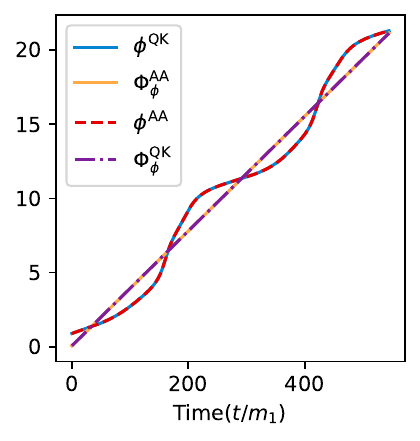}
		\caption{Azimuthal Phases}
		\label{fig:Azimuthal_Phases}
	\end{subfigure}
	\hfill
	\caption{The orbital phases for the first $500 m_1$ of a geodesic trajectory with orbital parameters $(a,p,e,x_I) = (0.9,7,0.5,0.75)$ and initial conditions $(\psi_0,\chi_0,\phi_0) = (2.7,1.8,0.9)$, produced by two independent methods. The QK angles $(\psi(t),\chi(t),\phi(t))$ oscillate with time, while the BLAAs evolve monotonically. By using the angle conversions developed in Sec.~\ref{section:QK_to_BL_AAs}, we see that the two methods agree to a high degree of accuracy.}
	\label{fig:Phase_Comparison_Plots}
\end{figure}

Numerically solving Eqs.~\eqref{eq:Quasi-Keplerian_EoM} with relative and absolute accuracy tolerances set to $10^{-13}$ takes $\sim 160$ ms and $\sim 4000$ time-steps. As seen in Fig.~\ref{fig:Phase_Comparison_Plots}, the numerical solver has to resolve the orbital oscillations in the solutions for the QK phases which requires small time-steps. By contrast, solving the AA system analytically is significantly faster (only $\sim 0.5$ ms). However, the conversion from AA to QK phases applied at each time-step takes $\sim 500$ ms, whereas converting from QK to AA phases at each time-step only takes $\sim 60$ms. So while it is faster to numerically solve the evolution equations for QK phases than to construct them from the analytic solutions for the AAs, for waveform production where speed is essential, it is the AA solutions that are more valuable.

\begin{figure}
	\begin{subfigure}[b]{0.49\textwidth}
		\centering
		\includegraphics[clip,width=\textwidth]{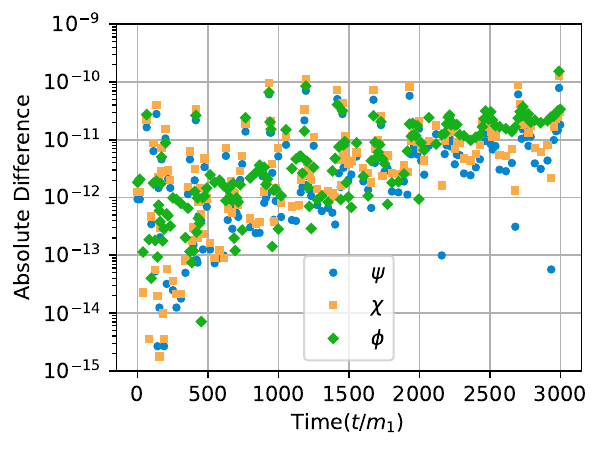}
		\caption{Difference in $(\psi(t),\chi(t),\phi(t))$}
		\label{fig:QK_Phase_Error}
	\end{subfigure}
	\hfill
	\begin{subfigure}[b]{0.49\textwidth}
		\centering
		\includegraphics[width=\textwidth]{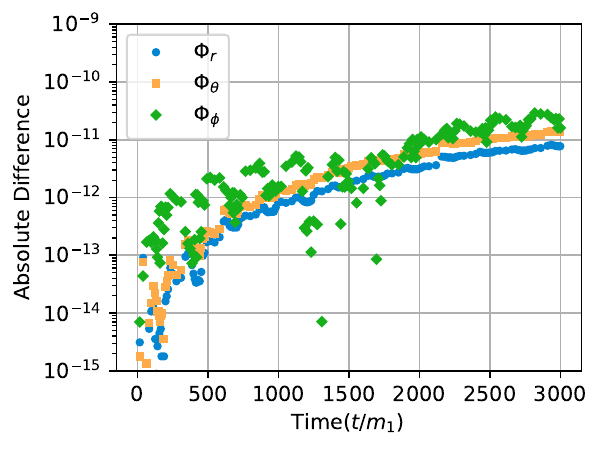}
		\caption{Difference in $(\Phi_r(t),\Phi_\theta(t),\Phi_\phi(t))$}
		\label{fig:Action_Angle_Error}
	\end{subfigure}
	\hfill
	\caption{The differences in the phases due to our two methods for solving for a geodesic trajectory. The QK angles have a time independent difference that is consistent with the tolerance of the numerical root-finder used in the conversion. The BLAAs have an difference that is consistent with the tolerances set on the numerical ODE solver which grows over time.}
	\label{fig:Phase_Error_Plots}
\end{figure}

The differences between the final solutions for both $(\psi(t),\chi(t),\phi(t))$ and $(\Phi_r(t),\Phi_\theta(t),\Phi_\phi(t))$ from the two methods is displayed in Fig.~\ref{fig:Phase_Error_Plots}. As we can see in Fig.~\ref{fig:QK_Phase_Error}, the conversion from AA to QK phases introduces a time independent error of $\lesssim 10^{-10}$ which is consistent with the error of the numerical solver of $10^{-10}$. The error on the action angles however is much smaller i.e. $\lesssim 10^{-11}$ as one does not need to numerically root-find. While the error is consistent with the tolerances on the numerical ODE solver, the error grows over time.\footnote{For error tolerances of $\lesssim 10^{-12}$, this growing numerical error from the ODE solver should still be a negligible source of error for an EMRI length trajectory of $\gtrsim 10^5 m_1$ especially compared to other systematic errors such as interpolation error and missing physics.} This means that for long enough trajectories, it would be more accurate to use the analytical solutions for the AAs which are exact and utilise the numeric conversion with a constant error, than to numerically solve for the QK phases and use the analytic conversion to AAs. 

\begin{figure} 
    \centering
    \includegraphics[clip,width=0.75\textwidth]{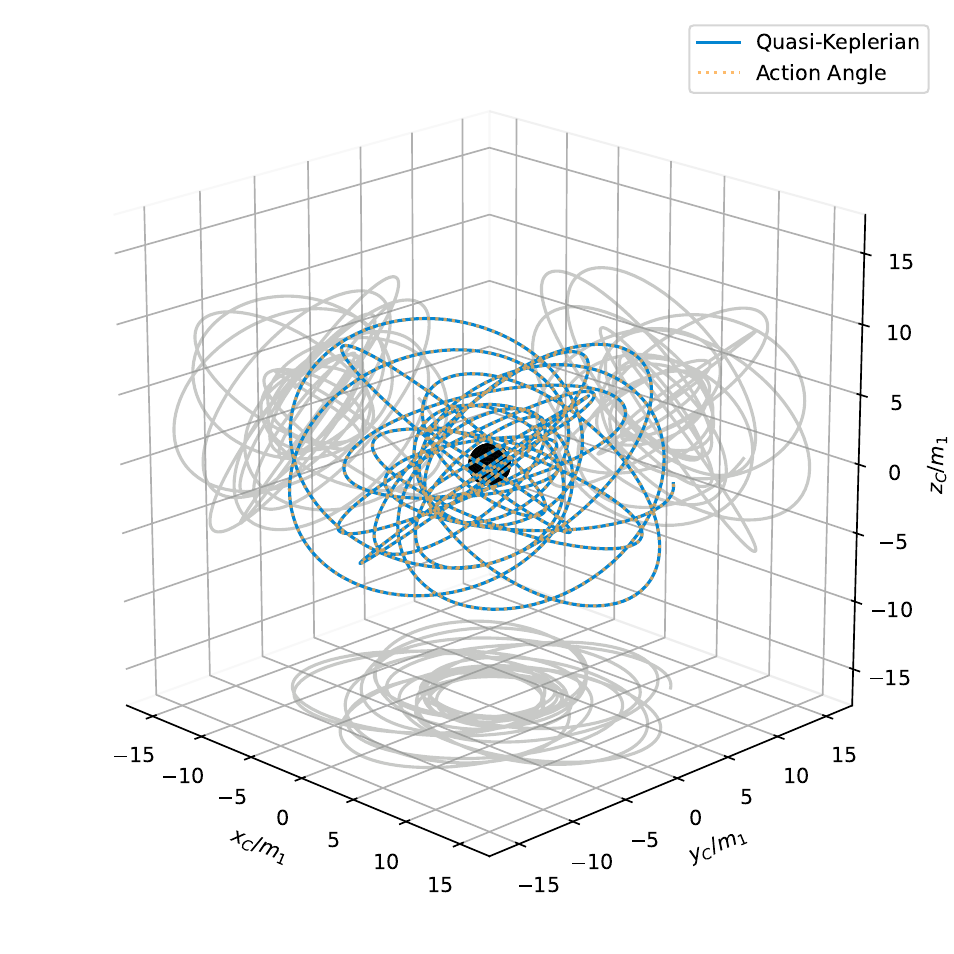}
    \caption{A visualisation of a geodesic trajectory with orbital parameters $(a,p,e,x_I) = (0.9,7,0.5,0.75)$ and initial conditions $(\psi_0,\chi_0,\phi_0) = (2.7,1.8,0.9)$ evolved for $3000 m_1$ using both the QK angles and the BLAAs.}
    \label{fig:Trajecotry}
\end{figure}

From both of these approaches, one can visualise the trajectory like in Fig.~\ref{fig:Trajecotry}. To do this, one must first obtain the solutions for $(\psi(t),\chi(t),\phi(t))$. From these, one can use Eqs.~\eqref{eq:Keplerian_Angles} along with $\theta = \arccos(z)$ to find solutions for the BL coordinates of the secondary with time $(r(t),\theta(t),\phi(t) )$. One can then construct Cartesian-like coordinates $(x_C(t),y_C(t),z_C(t))$ via the transformation:
\begin{subequations} \label{eq:Cartesian_Coordinates}
    \begin{gather}
			x_C(t) = r(t) \sin \theta(t) \cos \phi(t),   \quad  y_C  = r(t) \sin \theta(t) \sin \phi(t), \quad \text{and} \quad  z_C = r(t) \cos \theta(t).  \tag{\theequation a-c}
		\end{gather}
\end{subequations}
This allows us to create trajectory visualisations directly from the solutions for the BLAAs which agree with the results obtained via the QK angles. It also allows us to construct a vector pointing from the origin (centred on the primary) to the position of the secondary at a reference time $t_0$ by $\bm{n}(t_{0}) = (x_C(t_0), y_C(t_0), z_C(t_0))$. This is necessary to construct the source frame used by 0PA inspiral calculations \cite{lisa_rosetta_stone_2025}.

\section{Discussion}
In this note, we have presented a practical scheme for converting between the various sets of orbital angles used to describe both geodesic and inspiral trajectories in the Kerr spacetime. This includes an analytic conversion from QK angles $(\psi,\chi,\phi)$ to MTAAs $(q_r,q_z,q_\phi,q_t)$ and BLAAs $(\Phi_r,\Phi_\theta,\Phi_\phi)$ in terms of Jacobi elliptic integrals. While there is also an analytic conversion from MTAAs back to QK angles, there is no analytic conversion from BLAAs back to MTAAs. Instead we are forced to use an efficient root-finding scheme to invert this relationship. While more expensive and less accurate than the other analytic conversions presented here, this is still the most practical method to obtain MTAAs and QK angles starting from BLAAs. Utilising Eqs. \eqref{eq:Radial_Mino_to_QK}, \eqref{eq:Polar_Mino_to_QK}, \eqref{eq:Mino_to_BL}, and \eqref{eq:BL_to_Mino_Azimuthal}, one can relate the output of inspiral calculations that use BLAAs (such as the \texttt{FEW} package) to the BL coordinates of the secondary. Not only does this allow one to construct visualizations of the inspiral, but also allows for the construction of the source frame for 0PA inspirals. This can be converted to other frame choices as used by other GW source modelling approaches \cite{lisa_rosetta_stone_2025}.

We demonstrate the accuracy and computational cost of these conversions by computing a geodesic trajectory using QK angles and BLAAs and using the conversions presented here to ensure agreement between the two.
We find that one can invert the equations to within an absolute error in the phases of $\lesssim 10^{-10}$ in $\lesssim 1\,\text{ms}$. Such evaluation times are sufficiently quick for a single evaluation at the start of a trajectory calculation to ensure consistent initial conditions when comparing between different parameterisations to compute trajectories. 

This numerical conversion may become too costly to apply to every single time-step of a multi-year long EMRI trajectory for LISA data analysis. However, the main reasons for doing so are to ensure agreement between different trajectory calculations (as was done in the development of Refs.\cite{Lynch:2021ogr,Lynch:2023gpu,Drummond:2023wqc,Lynch:2024ohd}), or to produce trajectory visualisations. Neither of these use cases have the stringent runtime constraints of production level waveform calculations for LISA data analysis.

We have made our implementations available in the \texttt{KerrOrbitalAngleConversion} repository in Mathematica, C, and Python (via Cython) \cite{KerrAngleConversions}. We will integrate these conversions into tools already widely used by the community for ease of use. The Mathematica implementation will form the basis for an implementation in the \texttt{KerrGeodesics} package \cite{KerrGeodesics} which is part of the Black Hole Perturbation Toolkit \cite{BHPToolkit}. The C implementation with Cython bindings for Python will form the basis of an implementation in the \texttt{FEW} package.

\ack
This work makes use of the Black Hole Perturbation Toolkit. The authors thank Leor Barack and Adam Pound for their detailed feedback on an early version of this manuscript. OB is indebted to Leor Barack, Jonathan Gair, Lorenzo Speri, and Stas Babak for useful discussions related to this project. OB acknowledges financial support from the Grant UKRI972 awarded via the UK Space Agency.
\\
\bibliographystyle{IEEEtran}
\bibliography{AngleConversion}

\appendix{}

\section{Full expressions for $t_r$, $t_\theta$, $\phi_r$ and $\phi_\theta$} \label{section:Analytic_expressions_for_t_and_phi}
Here we give the full expressions for the oscillating functions $t_r$, $t_\theta$, $\phi_r$, and $\phi_\theta$ derived in Ref.~\cite{Fujita:2009bp} and simplified in Ref.~\cite{vandeMeent:2019cam}. First, we need to define the locations of the inner horizon $r_-$ and the outer horizon $r_+$ of the Kerr black hole which are given by
	\begin{equation}\label{eq:horizon_locations}
		r_\pm = M \pm \sqrt{M^2 - a^2}.
	\end{equation}
	This allows us to then define the following quantities
	\begin{subequations}
		\begin{gather}
			h_r \coloneqq \frac{r_a - r_p}{r_a - r_3} \quad \text{ and } \quad h_\pm \coloneqq h_r \frac{r_3 - r_\pm}{r_p - r_\pm}. \tag{\theequation a-b}
		\end{gather}
	\end{subequations}
	We then also define the quantities $\xi_r$ and $\xi_\theta$ via 
	\begin{subequations}
		\begin{gather}
			\xi_r \coloneqq \text{am}\left(\left.\frac{q_r K(k_r)}{\pi }\right |k_r\right) \quad \text{ and } \quad \xi_\theta \coloneqq \text{am}\left(\left.\frac{2 \left(q_\theta+\frac{\pi }{2}\right) K(k_\theta)}{\pi }\right|k_\theta\right). \tag{\theequation a-b}
		\end{gather}
	\end{subequations}
	These functions are in terms of elliptic integrals, including the incomplete elliptic integral of the second kind
    $E(\phi | m) \coloneqq  \int_0^\phi \sqrt{ 1 - m \sin^2 \theta } d\theta$,
	the complete elliptic integral of second kind $E(m) = E(\pi/2|m)$, the incomplete elliptic integral of the third kind
    $\Pi(n; \phi | m) \coloneqq  \int_0^\phi  d\theta/ \left( \left( 1- n \sin^2 \theta \right) \sqrt{ 1 - m \sin^2 \theta } \right)$,
	and the complete elliptic integral of the third kind $\Pi(n|m) = \Pi(n ; \pi/2|m)$.
    
	The radial oscillating terms are given by
	\begin{align}	
		\begin{split}
		t_r(q_r) &= -\frac{\En}{\sqrt{\left(1-\En^2\right) \left(r_a-r_3\right) \left(r_p-r_4\right)}} \Biggl[ 4 \left(r_p-r_3\right) \left(\frac{q_r}{\pi} \Pi \left(h_r|k_r\right)-\Pi \left(h_r;\xi_r|k_r\right)\right)
		\\&   -\frac{4 \left(r_p-r_3\right)}{r_+-r_-} \left(\frac{\left(r_+ \left(4-a \Lz/\En\right)-2 a^2\right) \left(\frac{q_r}{\pi} \Pi\left(h_+|k_r\right)-\Pi\left(h_+;\xi_r|k_r\right)\right)}{\left(r_p-r_+\right) \left(r_3-r_+\right)} - \left( + \leftrightarrow  -  \right) \right) 
		\\& +  \left(r_p-r_3\right) \left(r_a+r_p+r_3+r_4\right) \left(\frac{q_r}{\pi} \Pi \left(h_r|k_r\right)-\Pi \left(h_r;\xi_r|k_r\right)\right) 
		\\& + \left(r_a-r_3\right) \left(r_p-r_4\right) \left(\frac{h_r \sin \xi_r \cos \xi_r \sqrt{1-k_r \sin ^2\xi_r}}{1-h_r \sin^2 \xi_r }+\frac{q_r}{\pi} E\left(k_r\right)-E\left(\xi_r|k_r\right)\right) \Biggr],
		\end{split}\\
		\begin{split}
			\phi_r(q_r) &= \frac{2 a \En}{\left(r_+-r_-\right) \sqrt{\left(1-\En^2\right) \left(r_a-r_3\right) \left(r_p-r_4\right)}}  
			\\& \times \left[ \frac{\left(r_p-r_3\right) \left(2 r_+-a \Lz/\En\right) \left(\frac{q_r}{\pi} \Pi \left(h_+|k_r\right)-\Pi \left(h_+;\xi_r|k_r\right)\right)}{\left(r_p-r_+\right) \left(r_3-r_+\right)} - \left( + \leftrightarrow  - \right)\right] ,
		\end{split}
	\end{align}
	where we have adopted the shorthand $\left( + \leftrightarrow  -  \right)$ to indicate that one should repeat the previous term but with all positive symbols replaced with negative symbols (e.g., $r_+\leftrightarrow r_- $ etc.).
	The oscillating polar terms are given by
	\begin{subequations}
		\begin{align}
			\begin{split}
			t_\theta(q_\theta) &= \frac{\En z_+} {1-\En^2} \left(\frac{2}{\pi} \left(q_\theta+\frac{\pi }{2}\right) E\left(k_\theta\right)-E\left(\xi_\theta|k_\theta\right)\right),
			\end{split}\\
			\phi_\theta(q_\theta) &= - \frac{\Lz}{z_+}\left(\frac{2}{\pi} \left(q_\theta+\frac{\pi }{2}\right) \Pi \left(z_-^2|k_\theta\right)-\Pi \left(z_-^2;\xi_\theta|k_\theta\right)\right) .
		\end{align}
	\end{subequations}
\section{Quasi-Keplerian angles to Mino time action angles in the limiting cases}
In order to demonstrate that the QK angle to MTAA conversions are well behaved in their respective limiting cases, we have to derive the conversion in a different way.
Starting with Eq.~(\ref{eq:GeodesicEqsPandq}b) and using the chain rule along with Eqs.~\eqref{eq:Keplerian_Frequencies}, one obtains:
\begin{equation}
    \frac{d q_i}{d \lambda} = \frac{\partial q_i } {\partial \psi_i} \frac{d \psi_i}{ d \lambda} = \frac{\partial q_i } {\partial \psi_i} f_i  = \Upsilon_i,
\end{equation}
where $\psi_i = (\psi,\chi)$.
Rearranging the above and integrating with respect to $\psi_i$ gives us
\begin{equation} \label{eq:QK_to_Mino_Master}
    q_i =  \int_0 ^{\psi_i} \frac{\partial q_i }{\partial \psi_i} d \psi_i' = \int_0 ^{\psi_i} \frac{\Upsilon_i}{f_i(\psi_i')} d \psi_i'.
\end{equation}
Integrating Eq.~\eqref{eq:QK_to_Mino_Master} numerically recovers the same values as Eqs.~\eqref{eq:qr_of_psi} and \eqref{eq:qtheta_of_chi}.
\subsection{Radial conversion in the circular orbit limit} \label{section:Radial_Conv_Circular_Limit}
Obtaining an analytic solution to the radial component of Eq.~\eqref{eq:QK_to_Mino_Master} has proven illusive. For this analysis we only care about its behaviour as $e\rightarrow 0$. As such, we expand the integrand to first order in $e$ and obtain:
\begin{align}
\begin{split}
    q_r & = \int_0 ^{\psi} \left(1 +  \frac{\left(p (\mathring{r}_{3} + \mathring{r}_{4}) - 2 \mathring{r}_{3} \mathring{r}_{3} \right) e \cos \psi'}{2 (p - \mathring{r}_3) (p - \mathring{r}_4)} + \mathcal{O}(e^2) \right)d \psi' \\
    & = \psi +  \frac{\left(p (\mathring{r}_{3} + \mathring{r}_{4}) - 2 \mathring{r}_{3} \mathring{r}_{3} \right) e \sin \psi'}{2 (p - \mathring{r}_3) (p - \mathring{r}_4)} + \mathcal{O}(e^2),
\end{split}
\end{align}
where $\mathring{r}_3$ and $\mathring{r}_4$ denotes $r_3$ and $r_4$ evaluated at $e=0$. This is clearly well behaved in the $e \rightarrow 0$ limit and results in $\lim_{e \rightarrow 0} q_r = \psi$.

\subsection{Polar conversion in the equatorial orbit limit} \label{section:Polar_Conv_Equatorial_Limit}
The polar component of Eq.~\eqref{eq:QK_to_Mino_Master} does have an analytic solution in terms of Jacobi elliptic integrals given by
\begin{equation}
    q_\theta = \frac{\pi z_+ F(\chi | (\beta z_-^2/( \beta z_-^2 - z_+^2)))}{2 \sqrt{z_+^2 - \beta z_-^2} K(k_z)}
\end{equation}
This function has a well defined limit in the equatorial plane when $z_- \rightarrow 0$, which gives us: $\lim_{z_- \rightarrow 0} q_\theta = \chi$.

\end{document}